\documentclass[useAMS,usenatbib]{mn2e}

\usepackage{graphicx}
\usepackage{dcolumn}
\usepackage{amsmath}
\usepackage{textcomp}
\usepackage{amssymb}
\usepackage{color}

\newcommand{\emphr}[1]{{#1}}

\newcommand{\citee}[1]{[\cite{#1}]}

\newcommand{\VEC}[1]{\boldsymbol{#1}}
\newcommand{\rb}{\bar{r}}
\newcommand{\Hb}{\bar{H}}
\newcommand{\betab}{\bar{\beta}}
\newcommand{\omegaKo}{\bar{\Omega}_{\mathrm{K}}}
\newcommand{\omegabz}{\bar{\Omega}_{0}}
\newcommand{\xir}{\xi_{r}}

\newcommand{\qvb}{\VEC{q}}
\newcommand{\Jb}{\VEC{J}}
\newcommand{\Eb}{\VEC{E}}

\newcommand{\Tb}{T}
\newcommand{\Tbz}{\bar{T}_0}
\newcommand{\Tbu}{{T}_1}

\newcommand{\rhobz}{\bar{\rho}_0}
\newcommand{\rhobu}{{\rho}_1}

\newcommand{\pbz}{\bar{p}_0}
\newcommand{\pbu}{{p}_1}

\newcommand{\nbz}{\bar{n}_0}

\newcommand{\vb}{\VEC{v}}

\newcommand{\vbz}{\bar{\VEC{v}}_{0}}
\newcommand{\vbu}{{\VEC{v}}_{1}}

\newcommand{\vbuf}{{v}_{1\phi}}
\newcommand{\vbur}{{v}_{1r}}
\newcommand{\vbuz}{{v}_{1z}}

\newcommand{\Bb}{\VEC{B}}
\newcommand{\Bbz}{\bar{\VEC{B}}_{0}}
\newcommand{\Bbzs}{\bar{B}_{0}}
\newcommand{\Bbu}{{\VEC{B}}_{1}}
\newcommand{\Bbzz}{\bar{B}_{0z}}
\newcommand{\BbzzSQUARE}{\bar{B}^{2}_{0z}}
\newcommand{\Bbur}{{B}_{1r}}
\newcommand{\Bbuf}{{B}_{1\phi}}
\newcommand{\Bbuz}{{B}_{1z}}

\newcommand{\Ne}{\mathcal{N}}

\newcommand{\omegaK}{\Omega_{\mathrm{K}}}

\newcommand{\omegaT}{\Omega_{\mathrm{N}}}

\newcommand{\omegas}{\Omega_\mathrm{s}}
\newcommand{\omegaA}{\Omega_\mathrm{A}}
\newcommand{\ZA}{Z_\mathrm{A}}
\newcommand{\ZT}{Z_\mathrm{N}}

\newcommand{\ZR}{Z_\mathrm{R}}
\newcommand{\uA}{v_\mathrm{A}}
\newcommand{\vs}{v_\mathrm{s}}

\newcommand{\etav}{\eta_\mathrm{v}}
\newcommand{\presm}{p^\mathrm{m}_1}
\newcommand{\eref}[1]{Eq.(\ref{#1})}
\newcommand{\erefs}[1]{Eqs.(\ref{#1})}
\newcommand{\reff}[1]{(\ref{#1})}
\newcommand{\p}{\partial}

\title[Thermomagnetic instability of a rotating magnetized plasma disk]{Thermomagnetic instability of a rotating magnetized plasma disk}
\author[G. Montani, R. Benini, N. Carlevaro, A. Franco]{
Giovanni Montani$^{1,2}$\thanks{E-mail: giovanni.montani@frascati.enea.it},
Riccardo Benini$^2$\thanks{E-mail: riccardo.benini@gmail.com},
Nakia Carlevaro$^2$\thanks{E-mail: nakia.carlevaro@gmail.com},
Alessio Franco$^2$\thanks{E-mail: lexyalton@gmail.com}.\\
$^1$ ENEA - C.R. Frascati UTFUS-MAG, Rome (Italy).\\
$^2$ Physics Department, ``Sapienza'' University of Rome, P.le Aldo Moro 5, 00185 Roma (Italy).}
\begin{document}

\date{}

\maketitle

\label{firstpage}

\begin{abstract}
We analyze the stability of a thin plasma disk which is rotating around a compact astrophysical object and is embedded in the strong magnetic field of such a source. The aim of this study is the determination of a new type of unstable modes, able to replace the magneto-rotational instability profile for low $\beta$ values and for sufficiently small scales of the perturbations, where it fails. In particular, we consider the magneto-hydrodynamical scheme including a non-zero Nernst coefficient, corresponding to first-order collisional effects. As a result, modes with imaginary frequency lead to an instability regime when the magnetic tension vanishes. Finally, we show that, even in the presence of resistive effects, it remains a good candidate to ensure the onset of a turbulent behavior in the absence of the magneto-rotational instability.
\end{abstract}


\section{Introduction}
The description of the accretion process on a compact object constitutes one of the most relevant open questions in astrophysics, both for the understanding of crucial phenomena, like Gamma Ray Bursts \citee{Pi99} and Active Galactic Nuclei \citee{Kr99}, as well as because the accreting plasma can trigger the formation of matter jets, widely observed in nature \citee{Fe10}. The ``standard model'' for the formation and behavior of accretion disks is due to Shakura \citee{Sh73, SS73} and relies on the possibility to transport angular momentum outwards by means of the shear viscosity emerging from the disk differential rotation. Indeed, such a viscous effect can not be justified by the microscopic properties of the plasma and it is commonly accepted to be associated with the turbulent regime arising from the linear instability of the plasma dynamics with respect to small perturbations. The natural scenario able to induce turbulence is recognized in the Magnetorotational Instability (MRI) \citee{Ve59, Ch60} (for more recent developments, see \citee{BH91}), which holds as far as an arbitrarily small magnetic field is introduced in the problem.

The scope of the present analysis (referred to as a thin rotating disk) is to determine a new type of instability active in that region of the parameters where the MRI contribution is significantly suppressed, \emph{i.e.}, for low $\beta$ values and at small spatial scales. We succeeded in this attempt by fixing a Thermomagnetic Instability (TMI) which is triggered by a non-zero Nernst coefficient \emphr{(for a classical treatment of the thermomagnetic waves in plasma and TMI, see \citee{TS74, DU79, U81})}. Due to the microscopic value this term has in the quasi-ideal disk plasma, this instability will be relevant for very small scales where MRI is instead strongly suppressed. At small $\beta$ values, the only relevant dissipative effect corresponds to a non-zero plasma resistivity. We show how this additional contribution behaves as a damping of the instability growth rate, yet it is unable to suppress it by more than 50\%. This analysis is able to provide an alternative and complementary paradigm to MRI. 

\section{Basic features of the standard accretion model}\label{DUE}
Let us now briefly remind the standard accretion disk paradigm. In the Shakura model, the local equilibrium configuration \citee{BKL01} is mainly described by the hydrodynamical equations (for a critical approach to this, see \citee{MC12}). The radial force balance fixes the Keplerian nature of the disk angular frequency $\Omega$, while the vertical gravostatic equilibrium fixes the mass density profile and the half-depth of the disk $H$. Moreover, the continuity equation provides a constant value of the accretion rate $\dot{M}$ and the azimuthal force balance expresses the angular momentum transport across the disk in terms of the viscous stress tensor $\tau=-3\etav\Omega/2$, namely $\dot{M}(L-L^*)\sim-H\tau r^2$, where $L=\Omega r^2$ is the specific angular momentum ($r$ being the disk radial coordinate), $L^*$ is a given value of $L$ on the inner boundary layer of the disk and $\etav$ the shear viscosity coefficient. The viscous stress tensor corresponds to the relevant non-zero correlation function of the turbulent regime which, in the Shakura proposal, is responsible for the dissipative effects, \emph{i.e.}, $\tau=-\langle\rho v_r v_{\phi}\rangle\sim-\alpha\rho\vs^2$, where $\alpha$ is a parameter smaller than unity, $v_{r,\phi}$ denotes the radial and toroidal components of the velocity field, $\rho$ is the disk mass density and $\vs$ the sound speed. Comparing the expressions of the stress tensor, one easily gets $\etav\sim\alpha\rho\vs H$. The puzzle of such a scheme (which successfully accounts for many observed accretion features in astrophysical systems) consists of the stability that the hydrodynamical equilibrium manifests under small perturbations, preserving the axial symmetry. Therefore, unless peculiar scenarios (like the resonance phenomenon) are advocated, the paradigm of a turbulent regime would fall down.

The solution to this problem is offered by including an even small magnetic field in the equilibrium configuration, thus generating the MRI mentioned above. This scenario is well-grounded for astrophysical settings, because many accreting compact sources are endowed with a relevant intrinsic magnetic field and, for many real systems, it can be properly approximated by a dipole-like configuration that (for a thin disk) results in essentially a vertical field. In the MRI analysis, the local stability of a magnetically confined plasma disk can be then studied in a simplified perturbative scheme, in which the background is characterized by a purely Keplerian rotating configuration of constant mass density (and pressure) and embedded in a constant vertical magnetic field. The perturbations are consequently taken as local and incompressible (the so-called Boussinesq approximation) and propagating along the magnetic field direction. Furthermore, the linearity of the fluctuation dynamics allows us to use a plane wave form of wave-vector $k$ and frequency $\omega$. It can be shown that the normal modes associated to this scheme are characterized by the following dispersion relation (for a review, see \citee{BH98})
\begin{equation}
\omega^4-\omega^2[\Omega^2+2(k\uA)^2]+(k\uA)^2[(k\uA)^2-3\Omega^2]=0\;,
\label{mri}
\end{equation}
where $\uA$ is the Alfv\'en speed associated to the background magnetic field. \eref{mri} leads to the following condition for rising unstable modes $k<k_c\equiv\sqrt{3}\;\Omega/\uA$. MRI is therefore ensured in the presence of sufficiently small wavenumber (large scales). In this respect, it is worth underlining the presence of a minimum for the values of $k$ which corresponds to $k_{min}=\pi/H$ (\emph{i.e.}, to the maximum disk scale). Thus, the MRI mechanism is not applicable if $k_{min}>k_c$ (at sufficiently small scales) since all modes would result stable. The latter condition can be expressed in terms of the plasma $\beta$ parameter as $\beta<\pi^2/3$ and, since a large class of real accretion systems are indeed characterized by $\beta>\pi^2/3$, the corresponding MRI is commonly considered as the privileged mechanism for rising instabilities within an axisymmetric rotating plasma configuration. However, the existence of plasma disks for which $\beta$ is close to unity can not be excluded at all. In particular, in the case of very thin profile, MRI is not applicable as far as the magnetic field $B$ induced by the central object is sufficiently strong (we recall that $\beta\sim B^{-2}$). Since the sound speed is related to the disk local background temperature by the relation $\vs^2=5T/3m_\mathrm{i}$ ($m_\mathrm{i}$ being the ion mass), for a sufficiently cold and magnetized plasma disk we are led to search for new kind of instabilities, able to trigger the turbulent behavior.

\section{Thermomagnetic dynamical paradigm}\label{TRE}
We analyze the equilibrium configuration in the presence of a magnetic field strong enough to break down the MRI paradigm. In this respect, we consider the collisional effect described by the Nernst coefficient $\Ne$ \citee{LP81}, which emerges when the temperature gradient does not vanish. Such a transport coefficient is determined by the first-order correction to the equilibrium distribution function when the Boltzmann equation is expanded in powers of the synchrotron frequency inverse. In this case, \emphr{the dissipative energy density flux $\qvb$ is written as
\begin{align}
\qvb=\Ne\,\Tb\Bb\times\Jb\;,
\end{align}
where} $\Tb$, $\Bb$ and $\Jb$ are the temperature, the magnetic field and the current in the disk, respectively, and we use units such that $K_\mathrm{B}=1$. Correspondingly, the generalized Ohm law assumes the following form
\begin{align}\label{nernst}
\Eb+\vb\times\Bb/c=\Ne\Bb\times\nabla\Tb\;,
\end{align}
where $\Eb$ and $\vb$ denotes the electric and velocity field, respectively. \emphr{The microscopic expression of the Nernst coefficient reads as 
\begin{align}
\Ne=-\nu_\mathrm{ie}/[\sqrt{2\pi}\;c\,m_\mathrm{e}\,\omega_\mathrm{Be}^2]\;,
\end{align}
where} $\nu_\mathrm{ie}$ denotes the ion-electron collision frequency, $m_\mathrm{e}$ is the electron mass and $\omega_\mathrm{Be}$ the electron cyclotron frequency. 

Our analysis is implemented ([$r$, $\phi$, $z$] indicate cylindrical coordinates) at a fixed distance $r=\rb$ (in the following, we denote quantities evaluated at this radius with the bar symbol) from the central body of mass $M$ and the accretion disk is assumed to be thin, \emph{i.e.}, the half-depth $H(r)$ verifies the inequality $\Hb\ll\rb$. The local equilibrium configuration is described by a first-order perturbation of the usual magneto-hydrodynamics (MHD) equations and we denote the background quantities with $(...)_0$ and the fluctuations with $(...)_1$. A generic variable $A$ is thus perturbed near $\rb$ as $A=\bar{A}_0+A_1$, with $A_1\ll\bar{A}_0$. We underline that \emphr{the effects associated with the vector $\textbf{q}$} are neglected in the background dynamics. Since we are dealing with small scale perturbations, the local approximation results are well-grounded and the radial variation of the zeroth-order quantities can be effectively frozen to a given fiducial radius. 

The background configuration is specialized for an adiabatic equation of state $\pbz\sim\rhobz^{5/3}$, $\pbz(\rb,z)$ and $\rhobz(\rb,z)$ being the disk thermodynamical pressure and mass density, respectively. The zeroth-order radial momentum conservation (which reduces to the balance of the gravitational force with the centripetal one) locally fixes the Keplerian nature of the disk angular frequency $\Omega(r)$ as $\omegabz=\omegaKo=\sqrt{GM/\rb^3}$. \emphr{The background vertical local equilibrium determines instead the gravostatic profile of decay for the mass density (and for the pressure) as the vertical coordinate increases. Indeed, the equilibrium between the pressure and the vertical gravitational force simply results, in to the background profile \citee{BKL01}} $\rhobz(\rb,z)=\rhobz^*(\rb)(1-z^2/\Hb^2)^{3/2}$, where $\Hb=\sqrt{2\pbz^*/\rhobz^*\omegaKo^2}$ (pressure being $\pbz^*\sim(\rhobz^*)^{5/3}$ with $\rhobz^*=const.$). In our analysis, such a $z$-dependence is however disregarded in each background variables in view of the thinness of the disk ($|z|\leq\Hb\ll\rb$), thus reducing the mass density to $\rhobz=\rhobz^*(\rb)$, the pressure to $\pbz=\pbz^*$ and the temperature to $\Tbz=\Tbz^*(\rb)$, \emph{i.e.}, to assigned constants for the disk stability problem. Finally, the background magnetic and velocity fields read as $\bar{\VEC{B}}_0=(0,\,0,\,\Bbzz)$ and $\vbz=(0,\,\rb\omegaKo,\,0)$, respectively. These assumptions are well-grounded when studying the main stability properties of a real stellar accretion disk and they allow to simplify the analysis of the linear perturbation behavior by means of a Fourier expansion, as confirmed by their large employment in the literature \citee{BH98}.

\section{Thermomagnetic instability}\label{QUATTRO}
Addressing the perturbation theory, alternatively to MRI, we deal with a fluctuation dependence orthogonal to the background magnetic field, \emph{i.e.}, $\Bbz\cdot\nabla(...)=0$ (for an approach to TMI built up on the opposite limit, when the vertical gradient of the temperature dominates the equilibrium configuration of the disk and \emphr{when the energy transport is neglected by assuming that the energy is irradiated outwards the disk, \emph{i.e.}, \erefs{therm2} and \reff{therm3} are disregarded in the perturbation scheme}, see \citee{LMU10}). In this scheme, the magnetic induction equation, at the first perturbation order, stands as 
\begin{subequations}\label{filp}
\begin{align}
\p_t\Bbur=0\;,\\
\p_t\Bbuf+\omegaKo\,\Bbur=0\;,\label{filpr-phi}\\
\p_t\Bbuz+\Bbzz(\nabla\cdot\vbu+c\Ne\nabla^{2}\Tbu)=0\;,\label{filpz}
\end{align}
\end{subequations}
where, for the axial symmetry, we have neglected the $\phi$-derivative. While the magnetic tension vanishes identically, the magnetic pressure, at first order, reads as $\presm=\Bbz\cdot\Bbu/4\pi$ and plays a significant role in determining the disk stability properties.

The structure of the plasma velocity field leads us to write the Euler equations for the perturbative dynamics in the following form
\begin{subequations}\label{perturbeqs}
\begin{align}
\rhobz\p_t\vbur-2\rhobz\omegaKo\vbuf+\p_r(\pbu+\presm)=0\;,\label{perturbeqs1}\\
\rhobz\p_t\vbuz+\p_z(\pbu+\presm)=0\;,\label{perturbeqs2}\\
2\p_t\vbuf+\omegaKo\vbur=0\;.\label{perturbeqs3}
\end{align}
\end{subequations}
Correspondingly, the thermodynamics of the disk plasma is described by the evolution of the perturbed mass density $\rhobu$ and of the pressure $\pbu$, \emph{i.e.},
\begin{subequations}\label{therm}
\begin{align}
\p_t\rhobu+\rhobz\nabla\cdot\vbu=0\;,\label{therm1}\\
\p_t\pbu+\frac{5}{3}\,\pbz\nabla\cdot\vbu+
\frac{c}{6\pi}\,\Ne\Tbz\Bbz\cdot\nabla^{2}\Bbu=0\;.\label{therm2}
\end{align}
\end{subequations}

\eref{therm2} can be rewritten in terms of the temperature $\Tbu$, by using $\Tbz+\Tbu=(\pbz+\pbu)/(\nbz+n_1)$ (where $n$ denotes the number density), as
\begin{align}
\frac{3\nbz}{2}\,\p_t\Tbu+\pbz\nabla\cdot\vbu+
\frac{c}{4\pi}\,\Ne\Tbz\Bbz\cdot\nabla^{2}\Bbu=0\;,\label{therm3}
\end{align}
Finally, combining \eref{therm2} and \reff{therm3}, we easily obtain
\begin{equation}\label{combined23}
\p_t(\pbu/\pbz)+\nabla\cdot\vbu-\p_t(\Tbu/\Tbz)=0\;,
\end{equation}
and, deriving \eref{therm3} with respect to time using \erefs{filp} to express $\p_t\Bbu$, we get the following forth order equation for $\Tbu$ and $\vbu$
\begin{align}
\frac{3}{2}\frac{\p^2_{t}\Tbu}{\Tbz}-
\frac{2c\Ne\Tbz}{\betab}
\left[\nabla^{2}(\nabla\cdot\vbu)+c\Ne\nabla^4\Tbu\right]+\qquad\quad\nonumber\\
+\nabla\cdot(\p_t\vbu)=0\;,
\label{combeq}
\end{align}
where $\betab=8\pi\pbz/\BbzzSQUARE$.

If we now derive with respect to time \eref{perturbeqs1} using \eref{perturbeqs3} to eliminate $\p_t\vbuf$, we get the relation
\begin{equation}
\rhobz\p^2_{t}\vbur+\rhobz\omegaKo^2\vbur+\p_r\p_t(\pbu+\presm)=0\;,
\label{intermedeq}
\end{equation}
while the time derivative of the thermodynamical pressure is provided by \eref{combined23} and the corresponding magnetic term $\p_t\presm$ can be cast via \erefs{filp} as
\begin{equation}
\p_t\presm=\frac{1}{4\pi}
\Bbz\cdot\p_t\Bbu=-\rhobz\uA^2
(\nabla\cdot\vbu+c\Ne\nabla^{2}\Tbu)\;,
\label{magprestd}
\end{equation}
where the Alfv\'en velocity reads $\uA=\Bbz^{2}/4\pi\rhobz$.

Dealing with a vertical background magnetic field and working in axial symmetry, the request of an orthogonal perturbation dependence leads us to $\vbur=\vbur(t,r)$ and $\Tbu=\Tbu(t,r)$, only. In this scheme, we obtain the relation $\nabla\cdot\vbu=(1/r)\p_r(r\vbur)\simeq\p_r\vbur=\p_r(\p_t\xir)$, where we have used the small scale nature of the perturbations and $\VEC{\xi}=(\xir,\xi_{\phi},\xi_{z})$ denotes the plasma shift vector. Recalling that $\vs^2=5\pbz/3\rhobz$, in the adiabatic case, \eref{intermedeq} rewrites by means of $\xir$ as
\begin{align}
\p^3_{t}\xir+\omegaKo^2\p_t\xir-\Big[\frac{3}{5}\,\vs^2+\uA^2\Big]\p^2_{r}\p_t\xir+
\qquad\quad\nonumber\\
-\Big[c\Ne\uA^2+\frac{3}{5}\,\vs^2\p_r\p_t\Big(\frac{\Tbu}{\Tbz}\Big)\Big]=0\;.
\label{intermedeq2}
\end{align}
This equation must be coupled with \eref{combeq} expressed in terms of $\xi _r$. In view of the linear nature of such equations, we are able to consider the plane wave representation: $\xir=\xir^*\exp[i(kr-\omega t)]$ and $\Tbu=\Tbu^*\exp[i(kr-\omega t)]$, 
where $\xir^*$ and $\Tbu^*$ are constant values. Defining now $\omegas\equiv k\vs$, $\omegaA\equiv k\uA$ and $\omegaT\equiv-c\Ne k^2\Tbz$ (we recall that $\Ne<0$), Eqs.\reff{combeq} and \reff{intermedeq2} rewrites as
\begin{align}
\frac{\Tbu^*}{\Tbz}+\frac{4\omegaT+2i\betab\omega}{3\betab\omega^2+4\omegaT^2}\;
\omega k\xir^*=0\;,\label{combeqtr}\\
\omega^3-\big(\omegaKo^2+3\omegas^2/5+\omegaA^2\big)\omega-
\qquad\qquad\qquad\qquad\nonumber\\
-\big(\omegaA^2\omegaT+3i\omegas^2\omega/5\big)
\frac{\Tbu^*}{k\xir^*\Tbz}=0\;,\label{eqomega}
\end{align}
respectively. Combining the expression above, we finally get the following dispersion relation \begin{align}
\omega^4+\Big[\frac{4\omegaT^2}{3\betab}-\Omega^2\Big]\,\omega^2+
\frac{4}{3}\omegaA^2\omegaT\,i\omega+\nonumber\qquad\\
-\frac{4\omegaT^2}{3\betab}\Big[\Omega^2-\Big(1+\frac{\betab}{3}\Big)\omegaA^2\Big]=0\;,
\label{combdisprel} 
\end{align}
where we have set $\Omega^2\equiv\omegaKo^2+\omegas^2+\omegaA^2$ and we have used the adiabatic relation $\omegas^{2}=5\betab\omegaA^{2}/6$. Let us now divide \eref{combdisprel} for $\Omega^4$ and define the dimensionless quantities $Z_{\mathrm{A,N}}\equiv\Omega_{\mathrm{A,N}}/\Omega$ and $y=-i\omega/\Omega$. \eref{combdisprel} rewrites as
\begin{align}
y^4+\Big(1-\frac{4\ZT^2}{3\betab}\Big)\;y^2
-\frac{4}{3}\ZA^2\ZT\,y+\nonumber\qquad\\
-\frac{4\ZT^2}{3\betab}\Big[1-\Big(1+\frac{\betab}{3}\Big)\ZA^2\Big]=0\;.
\label{combdisprelComplex}
\end{align}
In this scheme, unstable modes clearly correspond to solutions with $\mathrm{Re}[y]>0$.

\begin{figure}
\begin{center}
\includegraphics[width=0.95\columnwidth]{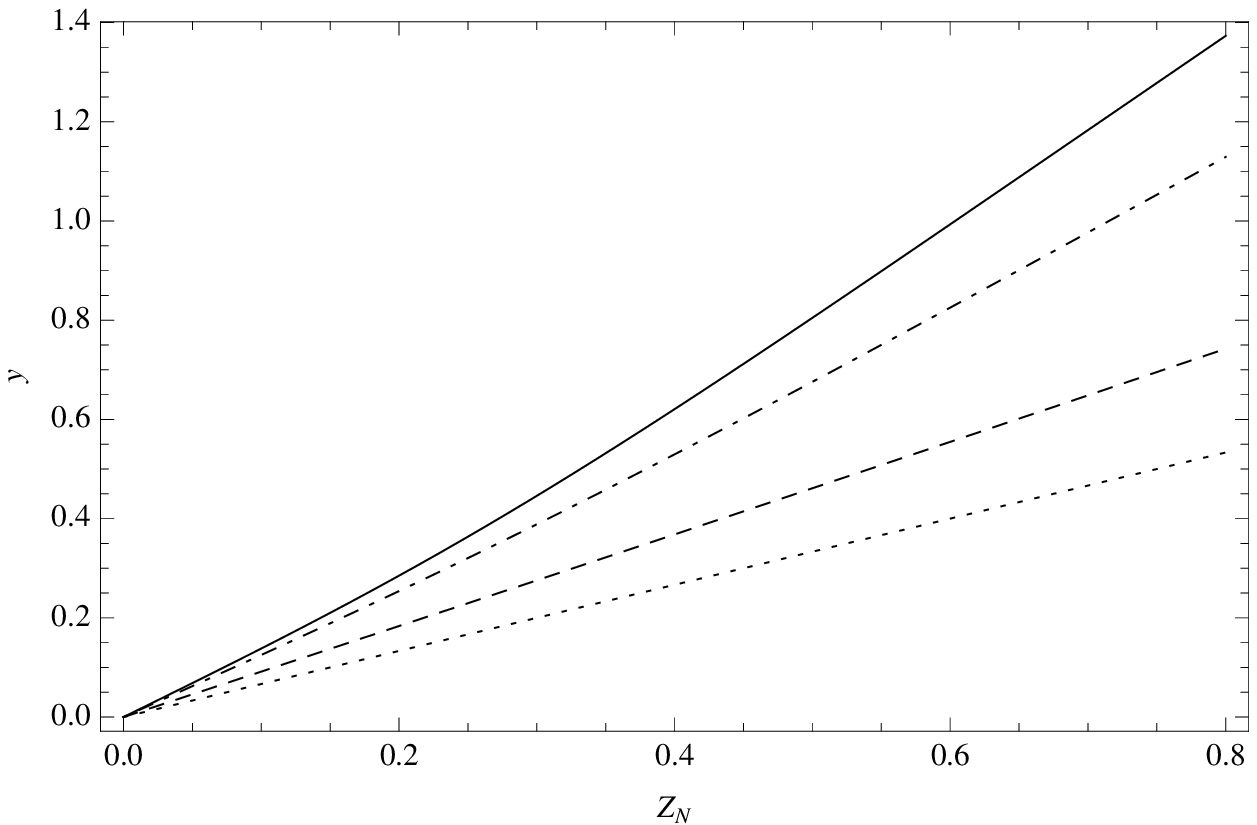}
\caption{Behavior of the positive solution (unstable mode) of \eref{combdisprelComplex} as a function of $\ZT$ for: $\betab=0.4$ (thick line), $\betab=0.6$ (dot-dashed), $\betab=1.5$ (dashed) and $\betab=3$ (dotted).}
\label{fig:yvszt}
\end{center}
\end{figure}

In the general case, there is always a mode fixed by the dispersion relation \reff{combdisprelComplex} that is unstable. This fourth-order algebraic equation admits two real solutions with positive and negative sign, respectively. There are other two solutions that can be complex conjugated or real depending on the values of the parameters; anyhow, the real part of such solutions is always greater than zero, resulting in two further unstable, in case oscillating, modes. In Figure \ref{fig:yvszt}, we plot the positive unstable solution which, in the limit of small $\ZT$, can be expressed as
\begin{equation}\label{eqlimite}
y=\tfrac{2}{3}\ZT\;\Big(\ZA^2+\sqrt{\left(1-\ZA^2\right)
\left(3-\betab\ZA^2\right)/\betab}\;\Big)\;.
\end{equation}
For ${\omegaKo}^2 \ll \omegas^2 + \omegaA^2$, we get $\ZA=6/(6+5 \betab)$ and we obtain the following growth rate $\gamma\equiv\Omega y$
\begin{equation}\label{solcomplexexp}
\gamma\simeq2\omegaT\Big(2+\sqrt{5}\sqrt{2+\betab}\;\Big)/\Big(6+5\betab\Big)\;.
\end{equation}
The considered limit is a very reasonable assumption already for a quite internal region of the disk. Let $R$ be the distance from the center in kilometer, for a temperature of $1$keV and for a perturbation scale of the order of $10$km, we obtain $\omegaKo/\omegas\sim10^{9/2}R^{-3/2}$ which is much smaller than unity for $R>10^5$.

Recalling that $\omegaT\equiv-c\Ne\Tbz k^2$ and using the expression of the Nernst coefficient, we can argue that
\begin{align}\label{stima}
\gamma&=\frac{ck^2r_\mathrm{c}}{\sqrt{2}\,\pi}\Big(\frac{m_\mathrm{e}c^2}{\Tbz}\Big)^{3/2}
\frac{2+\sqrt{5}\sqrt{2+\betab}}{6+5\betab}\;\betab\,\ell\simeq\nonumber\\
&\simeq 3.3\times 10^{-2}\,k_{5}^2\,T_4^{-3/2}\,\betab
\frac{2+\sqrt{5}\sqrt{2+\betab}}{6+5\betab}\;\;\mathrm{sec.^{-1}}\;.
\end{align}
Here, \emphr{$r_\mathrm{c}=e^{2}/m_e c^{2}\simeq2.8\times10^{-15}$m is the classical electron radius in Gaussian units}, $k_{5}$ the wave-number of the perturbation normalized to a length of $10^5$cm, $T_4$ the temperature divided by $10^4$K, and the Coulomb logarithm $\ell$ is set as $\ell=10$ \emphr{(we recall that the $\gamma$ parameter has been estimated by taking $\betab\sim1$)}. At sufficiently low temperatures and small scales of the disk plasma, such instability is relevant for the local behavior of the configuration profile. Indeed, in correspondence to a temperature $\Tbz$ range of $10^4-10^7$K (typical X-ray binary disks), and for $k_{5}=1$, we have $\gamma\simeq3\times10^{-2}-10^{-6}$sec.$^{-1}$. The growth rate of the obtained instability is relevant for the small time-scale evolution of the disk and it may concern the onset of a turbulent behavior in the configuration profile. However, the value of this evolution rate is at least two orders of magnitude below the MRI growth and therefore it is expected to be a higher order correction to the main unstable modes associated to the coupling of the differential disk rotation with the central object magnetic field, having a growth rate $\gamma\sim\omegaK$. Nonetheless, the relevance of TMI we fixed above consists essentially in its intrinsic different morphology with respect to  MRI, especially because it is associated to compressional unstable modes which can survive even in the region $\betab\sim1$, where MRI is instead strongly suppressed. In such a regime, the obtained instability is expected to be a dominant effect and it provides a reliable alternative scenario to MRI in determining a micro-scale turbulence. In other words, when $\betab\sim1$, TMI replaces MRI and in the outer regions of the disk its growth rate is comparable to the one MRI would have for higher $\betab$ values.

Concluding, we observe that, at very small spatial scales in the disk, for instance for characteristic wavelengths of the perturbations having the size of a meter, then such an effect strongly dominates the typical MRI behavior, becoming a really important feature in the disk micro-structure. Furthermore, considering such small scales is legitimated by the observation that the disk Debye length is always below the the millimeter scale. Thus, no questions can arise on the predictive nature of our approach at the scale of the meter.

\section{Resistive effects}\label{CINQUE}
It is a well-known feature that dissipative terms correspond to a general damping effect on the perturbation modes. In the following, we focus our attention on the electric and thermal resistivity contributions to the fluctuation profile. In this scheme, the generalized Ohm law \reff{nernst} and \emphr{dissipative energy density flux} rewrite as
\begin{align}\label{nernstres}
\Eb+\vb\times\Bb/c=\Jb/\sigma+\Ne\Bb\times\nabla\Tb\;,\\
\qvb=\Ne\,\Tb\Bb\times\Jb-\chi(\nabla T)\;,
\end{align}
respectively. Here, the coefficients $\sigma$ and $\chi$ stand for electric and thermal conductivities, respectively. We note that the temperature gradient retains only the radial component and, therefore, $\chi$ has to be intended as referred to the orthogonal direction. The approximate values of these parameters are given in \citee{LP81,BRA65}. Moreover, the perturbed magnetic induction and temperature equations allow us to estimate: $B_1/T_1=\sqrt{3\betab/4}\;\;\Tbz/\Bbzz$. This relation permits to evaluate the parameter range in which each single term is relevant with respect to the Nernst contribution in the perturbation scheme. Interestingly, the conditions for neglecting both the resistive and thermal conductivity effects lead indeed to restrictions on the $\betab$ parameter. In fact, it is easy to show that the thermal resistivity needs sub-thermal magnetic fields for it to contribute significantly to the system, \emph{i.e.}, $\betab\gg1$. Conversely, when $\betab\ll4/3$, dissipation is mainly due to the electric resistivity ($\sigma$ takes almost the same value in correspondence to the orthogonal and parallel direction of the magnetic field).

\begin{figure}
\begin{center}
\includegraphics[width=1\columnwidth]{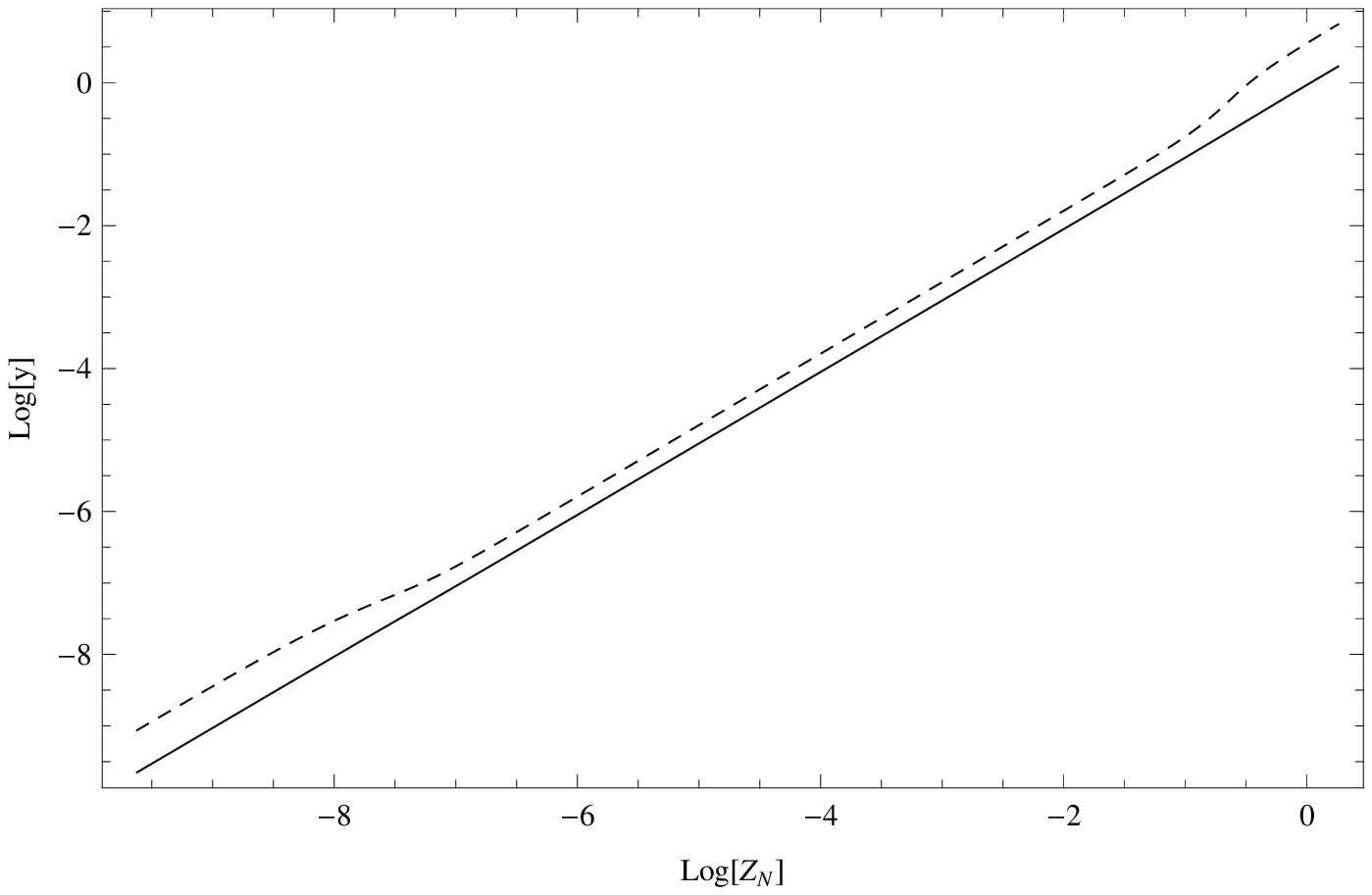}
\caption{Log-Log plot for $\betab=0.1$ of the unstable frequency $y$ as a function of $\ZT$ for the resistive case (thick line), as compared to the ideal case (dashed line).}
\label{fig:yvsztres}
\end{center}
\end{figure}

Since we are interested in analyzing the plasma behavior in the small-$\betab$ limit, we will consider the sole presence of electric conductivity. The derivation of the new dispersion relation follows the same steps we took for the ideal case. Using \eref{intermedeq}, and deriving with respect to time the axial component of the new induction equation,
\begin{align}\label{b1z}
\p_{t}^2\Bbuz+\Bbzz\p_t(\nabla\cdot\vbu&+c\Ne\nabla^{2}\Tbu)+\nonumber\\
&-c^2\,\p_{r}^2\p_t\Bbuz\,/\,4\pi\sigma=0\;,
\end{align}
we obtain a set of coupled differential equations in $\xi_{1r}$ and $B_{1z}$. It is easy to verify that the ensuing dispersion relation stands as 
\begin{align}\label{dispres}
y^4+\ZR\,y^3+\Big(1-\frac{4\ZT^2}{3\betab}\Big)\;y^2
+\Big[\ZR\Big(1-\ZA^2\Big)+\nonumber\quad\\
-\frac{4}{3}\ZA^2\ZT\Big]\,y-\frac{4\ZT^2}{3\betab}\Big[1-\Big(1+\frac{\betab}{3}\Big)\ZA^2\Big]=0\;.
\end{align}
where we introduced the quantity $\ZR=c^2k^{2}/4\pi\sigma\Omega$. The positive solution of the equation above is depicted in Figure \ref{fig:yvsztres}. The overall effect of electric resistivity is, as expected, a damping one of about the order of $50\%$ with respect to the ideal case.

\emphr{Finally, about the thermo-electromotive contribution to the Ohm law, we observe that, for perturbations propagating along the radial direction in a purely vertical field, the only surviving term is $\alpha_{\bot}(\nabla T)_{\bot}$ and it is easy to check that, in the considered parameter region, it can be neglected with respect to the Nernst term. In fact, the orthogonal thermo-electromotive coefficient takes the expression (see \citee{LP81, BRA65})
\begin{equation}
\alpha_{\bot}\simeq0.36
\Big(\frac{\nu_\mathrm{ie}}{\omega_\mathrm{Be}}\Big)^{2}\;.
\label{alfaort}
\end{equation}
The radial temperature gradient appears in both these two terms; thus, we have to compare $\alpha_{\bot}$ with respect to the modulus of the Nernst coefficient $\Ne$ times the background magnetic field intensity $\Bbzs$ which, by the definition of the Nernst coefficient, reads
\begin{equation}
|\Ne|\Bbzs\simeq\frac{1}{\sqrt{2\pi}\;e}
\Big(\frac{\nu_\mathrm{ie}}{\omega_\mathrm{Be}}\Big)\;,
\label{nb}
\end{equation}
or equivalently
\begin{equation}
|\Ne|\Bbzs\simeq \frac{T_{0}^{2}}{2\pi ce^{4}\ell\sqrt{n_0m_\mathrm{e}}}
\Big(\frac{\nu_\mathrm{ie}}{\omega_\mathrm{Be}}\Big)^{2}\;.
\label{nb1}
\end{equation}
Hence, it is easy to verify that $|\Ne|\Bbzs\gg\alpha_{\bot}$ holds true in typical accretion disks. As an example, considering a layer in which $T=10^{4}$K and $n_{0}=10^9$cm$^{-3}$, leads us to
\begin{equation}
|\Ne|\Bbzs\simeq1.16\times10^{10}\Big(\frac{\nu_\mathrm{ie}}{\omega_\mathrm{Be}}\Big)^{2}\;.
\end{equation} 
However, in the case a parallel thermo-electromotive contribution were present, \emph{i.e.}, $\alpha_{\|}(\nabla T)_{\|}$, the situation is reversed, in the sense that this effect would overcome the corresponding Nernst contribution. Such a statement follows directly from the expression of $\alpha_{\|}$, \emph{i.e.},
\begin{equation}
\alpha_{\|}\simeq\frac{1}{e}\Big(\frac{\mu_\mathrm{e}}{T}-4\Big)\;,
\label{alfapar}
\end{equation}
$\mu_\mathrm{e}$ being the electron chemical potential. Even when $\mu_\mathrm{e}\ll T$, this term clearly dominates the Nernst contribution \reff{nb} but is associated with the vertical gradient of the temperature, as far as the inequality $\nu_\mathrm{ie}\ll\omega_\mathrm{Be}$ is recognized, as for the present analysis. Nonetheless, this issue does not imply that the parallel thermo-electromotive effect would dominate also the Nernst term associated to the radial temperature gradient, when a vertical dependence of the perturbations is allowed. In fact, if we introduce such $z$-dependence, \emph{i.e.}, $\textbf{k}\cdot\textbf{r}=k_r r+k_z z$, it is easy to realize that our analysis holds as far as we have $k_z/k_r\ll\nu_\mathrm{ie}/\omega_\mathrm{Be}$. By other words, the contribution due to $\alpha_{\|}$ remains negligible as long as the vertical temperature gradient remains sufficiently smaller than the radial one.}

\section{Concluding Remarks}

Concerning the linear behavior of a thin rotating plasma disk, the present analysis has demonstrated the existence of a TMI, whose main feature is the striking complementarity to MRI, namely it is relevant at very small scales in the disk and also for low values of the $\betab$ parameter. The emergence of this instability is just a direct consequence of the kinetic properties of a plasma embedded in a sufficiently strong magnetic field. In fact, the obtained unstable normal modes are triggered by a non-zero Nernst coefficient, as fixed by a first-order expansion of the Boltzmann equation in inverse powers of the cyclotron frequency $\omega_\mathrm{Be}$. This effect is therefore present in any real disk for which such frequency is much larger than the ion-electron collision one, \emph{i.e.}, if $\omega_\mathrm{Be}\gg\nu_\mathrm{ie}$. In terms of the plasma parameters, this is equivalent to impose that $\Tbz^2/\sqrt{\betab\,\rhobz}\gg\;\sqrt{2\pi}\,ce^3\ell$. This condition is verified in a sufficiently hot plasma with a relatively low density, but in any case it stands for an enough small $\betab$ value. We stress that the constraint above is consistent with the present analysis. 

Despite the small value the Nernst coefficient takes in a typical accretion disk surrounding a compact astrophysical stellar object, TMI reaches growth rates even greater than the typical one of the MRI modes, as far as the scale of the linear fluctuations becomes of the order of the meter. This fact is even of large impact if we stress how, at such low spatial scales, MRI is completely suppressed. This new type of instability suggests that the small-scale evolution, especially in view of a possible turbulent behavior, can play a crucial role in the large-scale behavior too. In fact, the energy associated with the small-scale fluctuation due to TMI is comparable or greater than that present in the MRI on astrophysical size.

Furthermore, we have demonstrated that the TMI is, differently from MRI, present in that region of the thermodynamic and magnetic parameters of the disk, which corresponds to very small values of the $\betab$ parameter. In fact, although in this regime the resistivity damping can not be neglected, TMI is not affected for a factor greater than one-half and it conserves all its efficiency in destabilizing the small scale profile of the disk. Putting together all these considerations, we are naturally led to regard the instability here derived as a significant new effect in characterizing the main mechanisms for the angular momentum transport across the plasma disk configuration and the correspondingly accretion rate.

\section*{Acknowledgments}
This work was partially developed within the framework of the \emph{CGW Collaboration}
(www.cgwcollaboration.it).


\end{document}